\documentclass[nofootinbib]{revtex4-2}
\usepackage{graphicx,float}
\usepackage{amssymb}
\usepackage[normalem]{ulem}
\usepackage[mode=buildnew]{standalone}   
\usepackage{tikz}
\usepackage{rotating}

\newcommand{\be}{\begin{equation}} 
\newcommand{\ee}{\end{equation}}
\newcommand{\bc}{\begin{center}}
\newcommand{\ec}{\end{center}}

\usepackage[left=0.9in,right=0.9in,top=1in,bottom=0.9in]{geometry}

\begin{document}
\newcommand{\gs}[1]{\textcolor{red}{\it #1}}
\newcommand{\gsout}[1]{\textcolor{red}{\sout{#1}}}

\title{The ABC classification of exotic nuclei: a proposal}
\author{L. Fortunato$^{1,2}$\footnote{lorenzo.fortunato@pd.infn.it},  A. Vitturi$^{1,2}$\footnote{Posthumous author}}
\affiliation{Dip. Fisica e Astronomia ``G.Galilei", Universit\`a di Padova, via Marzolo 8, I-35131 Padova, Italy}   
\address{I.N.F.N. - Sezione di Padova}
\author{G. Singh$^3$\footnote{gagandeep.singh@ncbj.gov.pl}}
\address{National Centre for Nuclear Research, ul. Andrzeja So\l tana 7, 05-400 Otwock, Poland.}

\begin{abstract}
The large number of existing nuclear species and the long list of their possible different exotic properties, such as presence of a halo (A), Borromean structure (B), clusterization (C) and others, calls for a classification scheme that is universal, concise, categorical, informative, accessible and easily extensible. We provide here a first reasoned attempt to fill this gap with an abridged naming scheme, called $ABC$, based on definitions and properties that characterize modern nuclear physics. We limit our chart to light isotopes with $Z\le 10$ where most of these features appear presently.
\end{abstract}
\maketitle

The discoverer of the atomic nucleus, Ernest Rutherford, according to Ref. \cite{Birks}, once said, ``All science is either physics or stamp collecting" to underline the divide between sciences that are based on measurements that are to be understood by laws of mathematical character and sciences where mere enumerations of facts are  given or categorized without an effort to find an underlying mathematical description. While respecting the opinion of the very initiator of nuclear physics, systematic classifications and naming schemes are not without merit in science. Let us mention the taxonomy of Linnaeus \cite{Linn,Linn2} for the living organisms: even though it was mainly based on morphological features, most of it survived centuries of studies of physiology and biology, culminating with cladistic and DNA phylogeny of plants and animals. The system of classification of crystals, based on point group symmetries, is very useful because the naming scheme contains mathematical information so profound that, in fact, it has been imported from mineralogy to molecular physics. In astronomy, there are several classification systems \cite{Morg,Gray} for galaxies, stars, planets and any other object of investigation that can be used to categorize the innumerable variety of astronomical bodies into classes that share the same properties or behave in the same way. 

\begin{figure}[htbp]
		\centering
		\includegraphics[trim={0 0 0 0cm},clip,width=0.8 \columnwidth]{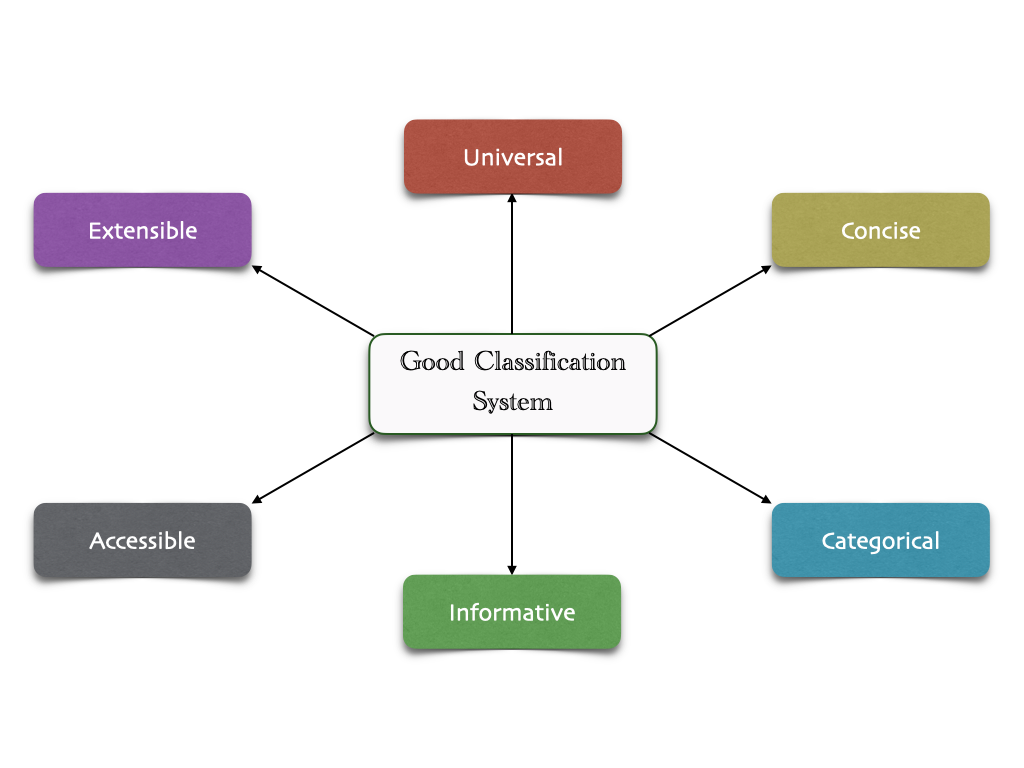}
		\caption{\label{fig: Class} Typical requirements of a good classification system.}
\end{figure}

The ancient philosophers gave a name to each substance encountered in nature, and speculated about the existence of four fundamental elements. The discoveries of chemists like Proust, Dalton, Gay-Lussac, Avogadro culminated in the celebrated periodic table of the elements of Mendeleyev. After Rutherford and Moseley (among others), a systematic investigation of all atomic species began and soon it was discovered that all elements, whose chemical properties are determined by the atomic number $Z$, i.e., the number of protons, might come in different guises, called isotopes, depending on the number of neutrons $N$. Aston, Soddy and Fayans are credited for having demonstrated the existence, the relevance and the properties of isotopes. Not all combinations of protons and neutrons might form a bound nucleus: at the moment of writing more than 3000 nuclear isotopes are known and the number is growing with the specialized progresses in acceleration and detection techniques all over the world \cite{Thoe}. Among them, only about 250-260 are stable, i.e., a small fraction of the total do not decay by emitting $\alpha, \beta$ or $\gamma$ radiation and they do not spontaneously fission, remaining as they are indefinitely, accounting for most of the ordinary matter. The others, on timescales that might range from billionths of a billionth of a second to times exceeding the life of the Universe, decay or transmute into the isotope of some other element. These nuclei are called unstable, or radioactive and are characterized by a number of unusual features. While there is a general rule as to when a collection of nucleons does constitute a nucleus \cite{Naz}, a traditional taxonomical categorisation is still lacking. 
We have to face a situation that natural phenomena offer a rich and varied spectrum of manifestations, and that any science that wants to describe them under the umbrella of a general law must face the problem of classification and naming. The most pressing problem, if anything, is to agree on conventions that are i) universal, ii) concise, iii) categorical, iv) informative, v) accessible and vi) easily extensible (in case future knowledge will require the addition of new traits). This is summarized in Fig. \ref{fig: Class}.
Modern nuclear physics is no exception to this need \cite{Thoe, Naz}. 
Of course, most measured properties, such as mass, charge, radii, spins, magnetic $M$1 and electric $E$2 moments are organized on the chart of nuclides and might be retrieved on digital databases with ease. The large number of species, though, and the richness of different phenomena, makes it very hard to collect all the information in a handy way. In other words, we have a problem at hand which lacks consistency due to our own incomplete knowledge in the field.

Thus, it was recently suggested at various seminars and meetings of nuclear physics in Italy that a classification scheme should be invented, much as the astronomers have for the galaxies, or the condensed matter physicists have for the types of  superconductors. This leads us to propose a scheme that meets the requirements discussed above, that we call the ABC classification of the properties of unstable nuclei. Each isotope should be labeled by the letters in alphabetical order corresponding to the features that it shows: 
\begin{itemize}
\item[A:]{stands for ``\textit{Alone}" (that is halo in Italian) with a subscript that counts the number of particles in the halo. For example, the two-neutron halo nucleus, $^6$He must be classified $A_2$, while $^8$He, with four neutrons in the halo must be classified $A_4$. The name has been introduced in nuclear physics in Ref. \cite{Jons87}.}
\item[B:]{stands for ``Borromean". A bound nuclear system is called Borromean, if it is composed of three subsystems, such that any binary subsystem is not bound. For example, returning to $^6$He, it has to carry a ``B" because it is thought as an $n+n+^4$He system and neither $^5$He, nor the di-neutron are bound. Since all Borromean nuclei discovered so far are halos, having a `$B$' in the classification scheme renders the `$A_2$' label to be implicitly present, but we wish to keep both names because they refer to different features.
The name has been introduced in nuclear physics from mathematics and knot theory in Ref. \cite{Zhuk}. }
\item[C:]{stands for ``Cluster". Several light nuclei are interpreted as being made of clusters (i.e., lumps of more than a single nucleon). For example, most of the properties of $^7$Li are well-understood in $\alpha + t$ models \cite{Lor}, or $^8$Be as $2\alpha$. If needed, a subscript might be used to specify the structure, but in most cases this requires a verbose description. We indicate that a nucleus $^Z_AX$ has a cluster structure if the lowest energy threshold is neither $^{A-1}_{Z-1}Y +p$ nor $^{A-1}_{Z}Y +n$. Sometimes a nucleus shows cluster features even if the lowest threshold is $p$ or $n$, such as $^{9}$Be.
The idea of clustering entered nuclear physics very early with the influential attempt by J. A. Wheeler to describe nuclei as nuclear molecules \cite{Whee}.}
\item[D:]{stands for ``Dripline". A neutron (or proton) drip-line is defined as the locus of points with null neutron (proton) separation energy, $S_n$ (or $S_p$) = 0. A drip-line nucleus marks the spot beyond which we do not expect nuclei to exist in a bound form.}
\end{itemize}
Additionally, we might use other letters to mark other important features: 
\begin{itemize}
\item[U:]{stands for ``Unbound". Some isotopes must be classified, because of their importance and they have been observed in experiments, even if they are unbound. For example $^8$Be is clearly a crucial nucleus in stellar nucleosynthesis even if it lives as a resonance for a very short time. Another good name for this category is ``Resonant" to highlight the transient nature of the system, before decay.}
\item[W:]{stands for ``Weakly-bound". The amount of binding energy determines several properties: most well-bound (traditional) nuclei have binding energies ranging from 7-9 MeV/A. Exotic nuclei, on the contrary, might have very low binding energies. The criterion to decide whether a nucleus is weakly- or strongly-bound is rather arbitrary and no net boundary can be easily identified. We propose to appose a `W' to the classification for species that have separation energies of the valence nucleon/cluster not exceeding $2.5$ MeV. By this definition, the deuteron, with binding energy of $\approx$ 2.224 MeV, $^6$He and $^7$Li are weakly-bound. This might seem arbitrary, but it has been chosen as a good compromise: for instance $^{16}$N is barely weakly-bound, while $^{12}$Be and $^{12}$B, with separation energies just above 3 MeV are not.}
\end{itemize}

By this classification, for instance, $^8$He is $A_4BDW$ since it could either be a Borromean system with the $n+n+^6$He configuration or have a tetraneutron skin in the $4n+^4$He configuration \cite{Duer22Nat}. Similarly, $^{11}$Li would be $A_2BDW$. The `$D$' occurs because the next two isotopes known, $^{12}$Li and $^{13}$Li are both unbound.
It may so happen that some of the properties of some nuclei might not be experimentally verified, thus, in keep with the traditional convention used for spin quantum numbers,  the letter for the corresponding property can be wrapped in parentheses. For example, the last bound isotope of Fluorine, $^{31}$F, is a proposed two-neutron halo, and hence, its Borromean property could be encoded using a $(B)$ instead of $B$.
Similarly some isotopes have been interpreted as possessing a cluster structure, and sometimes this can still be debated, or it might be that only some part of the spectrum shows cluster features, therefore some of the $C$ in our scheme might also be wrapped in parentheses, because there are indications that some clustering occurs even though the first energy threshold might correspond to the separation of a single nucleon.

\begin{sidewaysfigure}[htbp]
	\centering
\includegraphics[trim={0 0 0 0cm},clip,width=1.1\columnwidth]{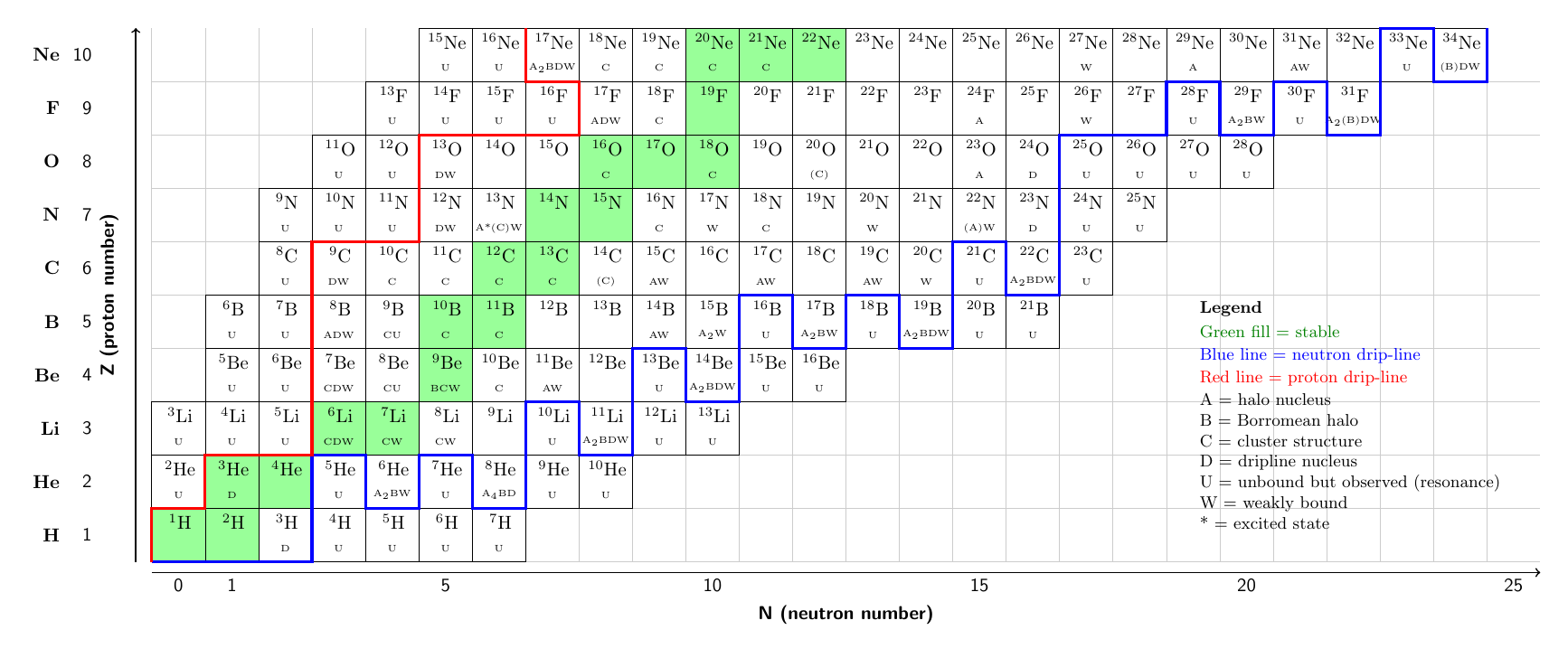}  
	\caption{The $ABC$ classification scheme applied to some of the light nuclei ($Z=1$--$10$).}
	\label{fig:abc-chart}
\end{sidewaysfigure}

Table \ref{Table} lists a few examples of how isotopes of an element can be classified using the proposed scheme. We show in   Fig. \ref{fig:abc-chart} a portion of the well-known Segr\'{e} chart, depicting the classification of most of the known nuclei up to $Z=10$ (as the drip-line has been confirmed only up to this number),  according to the ABC classification system. We have made the classification on the basis of a large number of measurements, datasets and references of which we give only a short list at the end of the bibliography section, namely \cite{Duer22Nat,NNDC,Till,TUNL,Iwa,JENDL,Oert}.

\begin{table}[!h]
	\begin{center}
		\begin{tabular}{|c|c|} \hline
	     $^5$He & U \\
	     $^6$He & BW \\
	     $^8$He & A$_4$BDW \\ \hline
	     $^6$Li & CSW \\
	     $^7$Li & CSW \\
	     $^{11}$Li & A$_2$BDW \\ \hline
	     $^8$Be & CU \\	\hline
         $^{31}$F & $A_{2}$(B)DW \\
	     \hline
		\end{tabular}
		\caption{\label{Table} Examples of ABCD classification.}
	\end{center}
\end{table}

We believe that this classification scheme shows all the marks, named above in points i)- vi), required to be a good classification scheme. Of course, the number and types of special attributes are not limited to the list that we have identified, and this classification scheme might be easily extended in the future by inventing further labels. 
One clear scope that this classification might achieve is that of highlighting patterns on the nuclear chart for further exploration.
With thousands of isotopes and many properties yet to be discovered, the nuclear chart contains inevitable classification gaps—much like Mendeleev’s original periodic table—that will eventually be filled as tools and techniques improve.

\section*{Data Availability} The manuscript has no associated data.

\end{document}